\documentclass[preprint,prd,showpacs,floatfix]{revtex4}
\usepackage{amsmath}
\usepackage{amsfonts}
\usepackage{amssymb}
\usepackage{bbm}
\usepackage{latexsym}
\usepackage[dvips]{graphicx}
\usepackage{graphics}
\usepackage{color}
\usepackage{multirow}

\usepackage{rotating}

\begin{document}

\title{Dark matter from Modified Friedmann Dynamics}
\author{Marek Szyd{\l}owski}
\email{uoszydlo@cyf-kr.edu.pl}
\affiliation{Astronomical Observatory, Jagiellonian University,
Orla 171, 30-244 Krak\'ow, Poland}
\affiliation{Complex Systems Research Centre, Jagiellonian University \\
Reymonta 4, 31-059 Krak{\'o}w, Poland}
\author{W{\l}odzimierz God{\l}owski}
\email{godlow@oa.uj.edu.pl}
\affiliation{Astronomical Observatory, Jagiellonian University,
Orla 171, 30-244 Krak\'ow, Poland}
\author{Jacek Golbiak}
\email{jgolbiak@kul.lublin.pl}
\affiliation{Department of Theoretical Physics, Catholic University of Lublin, 
Al. Rac{\l}awickie 14, 20-950 Lublin, Poland}

\date{\today}

\begin{abstract}
The contemporary cosmic expansion is considered in the context of Modified
Friedmann Dynamics (MOFD). We discuss some relativistic model exploring analogy
to MOND modification of Newtonian dynamics.
We argue that MOFD cosmologies can explain fraction
of dark matter in the accelerating Universe. We discuss some observational
constraints on possible evolutional MOFD scenarios of cosmological models
coming from SN Ia distant supernovae. We show that Modified Newtonian Dynamics
can be obtained as a Newtonian limit of more general relativistic models.
with polytropic component of Equation of State. They
constitute a special subclass of generalized Cardassian models basing on
generalization of the Raychaudhuri equation rather than on generalization of
the Friedmann first integral. We demonstrate that MOND cosmologies are compatible
with observed accelerated phase of expansion of current universe only for
high value of cosmological constant. The Bayesian framework of model selection
favored this model over $\Lambda$CDM model if $\Omega_{m,o}$ is fixed but this
evidence is not significant. Moreover obtained from statistical analysis value 
of the MOND characteristic $\beta$ parameter is far from value required for 
explanation of the flat rotation curves of spiral galaxies.
 
\end{abstract}
 
\pacs{98.80.Bp, 98.80.Cq, 11.25.-w}
 
\maketitle

\section{Introduction}
 
The idea that dark matter manifestation in flat rotation curves of spiral
galaxies \cite{Milgrom:1983} is a consequence of Modified Newtonian Dynamics
(MOND) seems to be both intriguing and controversial how claims Lue and
Starkman \cite{Lue:2003ky,Lue:2003if}. In principle there is a simple way to
test this theory \cite{Milgrom:1983} by consideration observational consequences
coming from cosmological models basing on these modifications of gravitational
interactions at the late time. Of course the MOND theory is Newtonian but it
should be obtained as a limit of more general and fundamental theory (Modified 
General Relativity -- MOGR). If we consider homogeneous and isotropic cosmology
in this theory then the modified Friedmann equations will describe the evolution
of the Universe.
 
Recently, different modifications of Friedmann equation were proposed
\cite{Freese:2002sq}. Motivation of this model is to explain the current
acceleration of the Universe \cite{Perlmutter:1998np,Riess:1998cb} without
references to the unknown form of dark energy. In this scenario (called the
Cardassian or polytropic expansion) there is no dark energy component but the 
universe is matter dominated. It is accelerating due to the adding a certain 
additional term to Friedmann
first integral of Einstein equations (with the Robertson-Walker symmetry).
An important question is whether there is any connections between the MOND
driven cosmology and the Cardassian one.
 
In the paper by Lue and Starkman \cite{Lue:2003ky} it is presented interesting 
idea of derivation modified theory of gravity from constraint coming from the
fundamental Birkhoff law. In this approach authors explain how the cosmic
acceleration is generated through these modifications.
 
In this paper we incorporate MOND for the late time cosmological scenario, while
early stages of evolution are dominated by usual matter  described in a
standard way by general relativity. For simplicity (without losing the generality)
we assume that our universe is flat. In derivations of basic dynamical
equations both for Newtonian and relativistic model we use particle-like
description \cite{Szydlowski:2003cf}. In this approach the evolution of the
universe is represented by a motion of a unit mass particle under the action
of a one-dimensional potential $V(a)$ which can be simply obtained from the MOND
gravitational acceleration postulate. The position variable a is a 
scale factor of the universe and all dynamics is determined by the
potential function through the analog of Newtonian equations.
 
The Cardassian models base on the generalization of the Friedmann first
integral by adding in r.h.s. a term which is called the Cardassian term, i.e.,
\begin{equation}
H^{2}=\frac{\rho}{3}+B\rho^{n}-\frac{k}{a^{2}},
\label{eq:1}
\end{equation}
where $\rho$ is the energy density. If the source of gravity is a perfect
fluid with pressure $p=\gamma\rho$ ($\gamma = \text{const}$) then
$\rho=\rho_{0}(a/a_{0})^{-3(1+\gamma)}$, $a_{0}$ is a present value of the
scale factor, $k=0,\pm 1$ is the curvature constant, $H=(\ln{a}){}\dot{}$ is
the Hubble function. Note that equation~(\ref{eq:1}) is the first integral
of the generalized Einstein equation for the Robertson-Walker symmetry. The
basic equations constitute the system
\begin{subequations}
\label{eq:2}
\begin{align}
\dot{H} &= -H^{2}-\frac{\rho}{6}\left(1+3\gamma\right)
+\frac{B\rho^{n}}{2}, \label{eq:2a}\\
\dot{\rho} &= -3H\rho(1+\gamma) \label{eq:2b}
\end{align}
\end{subequations}
with the first integral in the form
\begin{equation}
\left(\frac{H}{H_{0}}\right)^{2}=\Omega_{\gamma,0}\left(\frac{a}{a_{0}}\right)^{-3(1+\gamma)}
+\Omega_{\text{Card},0}\left(\frac{a}{a_{0}}\right)^{-3n(1+\gamma)}
+\Omega_{k,0}\left(\frac{a}{a_{0}}\right)^{-2}.
\label{eq:2c}
\end{equation}
Equation (\ref{eq:2a}) is called the Raychaudhuri equation, while equation
(\ref{eq:2b}) is the conservation equation. It seems to be more natural to
generalize the Raychaudhuri equation instead of its first integral. In the
last case we obtain a more general theory containing MOND cosmologies as
a special case. If we consider the standard Cardassian models, then in
the right-hand side of equation \ref{eq:2c}) only power low terms can of type 
$a^\beta$ can appears while in the MOND cosmologies some part of potential 
is logarithmic type.

Our basic idea is to explain the fraction of dark matter in the Universe in
analogy to the Milgrom \cite{Milgrom:1983} explanation of flat rotation 
curves of spiral galaxies, i.e. in terms of the MOND conception rather than 
mysterious dark energy.
 
The organization of our paper is as follows. In section~2 we provide a brief
summary of the features of the Cardassian models and generalized Cardassian
models that are relevant for our further discussion. The particle-like
description of MOFD cosmologies and constraining model parameters in the
light of SNIa data (based on the Riess sample) are presented in section~3.
Finally in section~4 some concluding remarks and perspectives for analysis
of cosmology in the new MOFD (or MOGR) paradigm are formulated.

\section{Generalized Cardassian models as a natural generalization of
FRW models.}
 
By the generalized FRW Cardassian models we understand models which dynamics
is governed by the generalized Raychaudhuri equation and conservation condition
\begin{subequations}
\label{eq:3}
\begin{align}
\frac{\ddot{a}}{a} &= -\frac{\rho(a)}{2}\left(\frac{1}{3}+\gamma(a)\right)
-\frac{f(\rho(a))}{6}, \label{eq:3a} \\
\dot{\rho} &= -3\left(\frac{\dot{a}}{a}\right)
(1+\gamma(a))\rho(a) \label{eq:3b},
\end{align}
\end{subequations}
where a dot denotes differentiation with respect to the cosmological time
$t$, $f(\rho(a))$ defines the type of modification of the standard Raychaudhuri equation 
for FRW cosmology which holds for $f=0$.
 
System (\ref{eq:3}) has a first integral in the form
\begin{equation}
\rho_{\text{eff}}-3\frac{\dot{a}^{2}}{a^{2}}=3\frac{k}{a^{2}},
\label{eq:4}
\end{equation}
where $\rho_{\text{eff}}(a)$ plays the role of effective energy density
(see Appendix). Equation (\ref{eq:4}) is independent (directly) on the
special form of matter the Universe is filled with. In the generic case if
we put into (\ref{eq:3a}) $f(\rho)\propto \rho^{n}$ and $\gamma(a)$ like for
a mixture of noninteracting matter and radiation then the usual class of the
Cardassian models is recovered. However let us note that (\ref{eq:4}) with
$\rho_{\text{eff}}(a)=\rho(a)+3B\rho^{n}$ does not play the role of the first
integral in the special case when $f(\rho)a\propto a^{-1}$. It is just the
case of the MOND cosmologies. To illustrate this let us consider the simplest
case of single fluid with energy density $\rho$ and $\gamma=\text{const}$.
Then from equation (\ref{eq:3b}) we obtain
\begin{equation}
\rho=\rho_{0}\left(\frac{a}{a_{0}}\right)^{-3(1+\gamma)}
\label{eq:5a}
\end{equation}
or in term of density parameter $\Omega_{i,0}\equiv\rho_{i,0}/3H_{0}^{2}$
\begin{equation}
\Omega_{i}=\Omega_{i,0}\left(\frac{a}{a_{0}}\right)^{-3(1+\gamma)}
\label{eq:5b}
\end{equation}
Let us substitute $f(\rho)=3B\rho^{n}$. Hence (\ref{eq:3a}) assumes the form
\begin{equation}
\frac{\ddot{a}}{a}=-\frac{\rho_{0}}{2}\left(\frac{1}{3}+\gamma\right)
\left(\frac{a}{a_{0}}\right)^{-3(1+\gamma)}-\frac{\rho_{0}^{n}B}{2}
\left(\frac{a}{a_{0}}\right)^{-3n(1+\gamma)}
\label{eq:6}
\end{equation}
It would be useful to consider in (\ref{eq:6}) two cases
\[
n\neq\frac{2}{3(1+\gamma)} \quad \text{or} \quad n=\frac{2}{3(1+\gamma)}.
\]
In the first case we obtain a class of cosmologies called the Cardassian models.
They can be treated as standard cosmological models where the universe is
filled with a mixture of non-interacting perfect fluids with the equation of state
$p=\gamma \rho$ for the first and $p=[n(1+\gamma)-1]\rho=w\rho$ for the second
one. Therefore the Cardassian models with a single fluid have dynamics in the
form of a two-dimensional dynamical system
\begin{align}
\dot{x} &= y \\
\dot{y} &= -\frac{1}{2}\left\{\Omega_{\gamma,0}x^{-2-3\gamma}(1+3\gamma)
+\Omega_{\text{Card},0}x^{-2-3w}(1+3w)\right\} \equiv-\partial V/\partial x,
\label{eq:8}
\end{align}
where $\Omega_{\gamma,0}=\rho_{\gamma,0}/3H_{0}^{2}$,
$\Omega_{\text{Card},0}=B\rho_{\gamma,0}^{n}/H_{0}^{2}(1+3w)$, are density
parameters of matter and the fictitious Cardassian fluid respectively,
$x=a/a_{0}$ is a dimensionless scale factor in the units of its present value
$a_{0}$. A dot here denotes differentiation with respect to re-parameterized
time variable $\tau$ defined as
$t\rightarrow\tau \colon dt \left|H_{0}\right|=d\tau$.
Of course $\Omega_{\gamma,0}+\Omega_{\text{Card},0}+\Omega_{k,0}=1$ is satisfied.
Note also that the potential function $V$ is determined modulo to any additive
constant.
 
Because $n\not=\frac{2}{3(1+\gamma)}$ system (\ref{eq:8}) possesses the
first integral in the form
\begin{equation}
\frac{y^{2}}{2}+V(x) \equiv 0,
\label{eq:9}
\end{equation}
where $V(x)=-\frac{1}{2}\left\{\Omega_{\gamma,0}x^{-1-3\gamma}
+\Omega_{\text{Card},0}x^{-1-3w}+\Omega_{k,0}\right\}$ and the constant in
$V$ should be chosen such that $\sum_{i}\Omega_{i,0}=1$. The constraint
condition $\sum_{i}\Omega_{i,0}=1$ in the general relativity reveals the fact
that both matter and geometrical term are related. The fact that
$\Omega_{\text{Card},0}$ does not contribute in this relation is a reflection
of the fact that we are beyond the standard cosmology. Note that it can be
estimated only from the observations.
 
Let us comment now the second case of $n=2/3 (1+\gamma)$. Then the potential 
function assumes very special form with logarithmic component:
\begin{equation}
V(x)=-\frac{1}{2}\left\{\Omega_{\gamma,0}x^{-1-3\gamma}
+\Omega_{C,0}\ln{x}+\Omega_{k,0}+(1-\Omega_{\gamma,0})\right\},
\label{eq:10}
\end{equation}
where $\Omega_{k,0}^{\text{eff}}=\Omega_{k,0}+(1-\Omega_{\gamma,0})$. Then
if we substitute this form into the(\ref{eq:9}) we can obtain the form of the
first integral for this case (see Appendix). Note that both last two terms in 
(\ref{eq:10}) of the same type can be defined in one term which we called 
effective curvature density parameter. Usually the form of first integral (\ref{eq:9}) is treated as a
starting point to further analysis of the generalized Friedmann equation.
 In our opinion the generalization of FRW equations on the level
of the Raychaudhuri equation seems to be methodologically more correct
procedure than generalization of its first integral. Moreover is more general 
because one additional case is included.
 
Finally the generalized Cardassian models in our terminology constitute
larger class of models and both cases for which both $n\neq\frac{2}{3(1+\gamma)}$
and $n=\frac{2}{3(1+\gamma)}$ belongs to this class. There are two parameters
characterizing models of this class ($n$, $\Omega_{\gamma,0}$) if
$\Omega_{k,0}=0$.

\section{Particle-like description of MOND and MOFD cosmologies.}
 
In MOND the gravitational acceleration $g$ exerted by a body of mass $M$ at
the radial distance $a$ obeys the relationship
\begin{equation}
g \propto \left\{ \begin{array}{lll}
 -a^{-2} & \quad \text{for} \quad & \left|g\right| > g_{0}\\
 -a^{-1} & \quad \text{for} \quad & \left|g\right| < g_{0}\\
\end{array} \right.
\label{eq:11}
\end{equation}
where $g_{0}$ is a critical value of acceleration. Hence the potential of
the gravitational field can be simply calculated from the formula
\begin{equation}
V(a)=-\frac{1}{M}\int_{0}^{a}g(a)da \propto
\left\{
\begin{array}{lll}
-a^{-1} & \quad \text{for} \quad & \left|g\right| > g_{0}\\
\ln{a}  & \quad \text{for} \quad & \left|g\right| < g_{0}\\
\end{array}
\right.
\end{equation}
At first we can build the Newtonian (\ref{eq:11}) cosmological models basing
on the particle-like description of quintessential cosmology developed by us
earlier \cite{Szydlowski:2003cf}. Following this approach the dynamics of
Newtonian cosmological models can be represented by a motion of the
particle-universe under the action of a one-dimensional potential $V=V(a)$,
$a$ is a scale factor of the universe plays the role of positional variable.

The heuristic method of obtaining Newtonian modified potential is basing on
consideration Schwarzchild solution of relativistic model. We start from 
Newtonian model potential and then derive relativistic model.
In the FRW cosmology the evolution of the universe can be derived from
the Hamiltonian which in terms of dimensionless variable takes the form
\begin{equation}
\mathcal{H}=\frac{1}{2}y^{2}+V(x) \equiv 0,
\label{eq:13}
\end{equation}
where $x=a/a_{0}, y=\dot{x}$ and $V(x)$ is in the form
\begin{equation}
V(x)=-\frac{1}{2}\left\{\Omega_{\gamma,0}x^{-1-3\gamma}
+\Omega_{\text{MOND},0}\ln{x}+(1-\Omega_{\gamma,0})\right\}.
\label{eq:14}
\end{equation}
 The last term $(1-\Omega_{\gamma,0})$ in (\ref{eq:14}) plays only
the role of negative curvature term
$\Omega_{k,0}^{\text{eff}}=1-\Omega_{\gamma,0}$. The Hamiltonian is defined
on zero energy level $\mathcal{H}=E=0$. The motion in the configuration
space is defined in the domain admissible for motion:
\begin{equation}
\mathcal{D}_{0}=\{x \colon V(x)\leq0\}
\end{equation}
From (\ref{eq:13}) and (\ref{eq:14}) we obtain the counterpart of the
Friedmann equation in our theory. Of course if we substitute $H=H_{0}$ and
$x=1$ then we recover $\sum_{i}\Omega_{i,0}=1$ as a constraint on density
parameters from relation
\begin{equation}
H^{2}(x)=H_{0}^{2}\left\{\Omega_{\gamma,0}x^{-3(1+\gamma)}
+\Omega_{\text{MOND},0}x^{-2}\ln{x}+\Omega_{k,0}^{\text{eff}}x^{-2}\right\}
\label{eq:15}
\end{equation}
or in the terms of redshift
\begin{equation}
H^{2}(z)=H_{0}^{2}\left\{\Omega_{\gamma,0}(1+z)^{3(1+\gamma)}
-\Omega_{\text{MOND},0}(1+z)^{2}\ln{(1+z)}
+\Omega_{k,0}^{\text{eff}}(1+z)^{2}\right\}.
\end{equation}
By comparing (\ref{eq:15}) with (\ref{eq:10}) we find strictly correspondence
between a special second class of the Cardassian models with
$n=\frac{2}{3(1+\gamma)}$ and MOND cosmologies.
 
From the Newtonian analogue of the equation of motion
$\ddot{x}=-\partial V/\partial x$ we find that
\begin{equation}
\ddot{x}=\frac{1}{2}\left\{\Omega_{\gamma,0}(-1-3\gamma)x^{-2-3\gamma}
+\Omega_{{\rm MOND},0}x^{-1}\right\}
\end{equation}
The universe is accelerating at the present epoch (x=1) if only
\begin{equation}
\Omega_{{\rm MOND},0}>(1+3\gamma)\Omega_{\gamma,0}
\end{equation}
Therefore for $\gamma=0$ (dust) $\Omega_{\text{MOND},0}>\Omega_{m,0}$ is
required if $\Omega_{k,0}^{\text{eff}}=0$.
 
The values of model parameter ($\Omega_{m,0}$, $\Omega_{\text{MOND},0}$) can
be obtained from the fitting procedure to SNIa data. The luminosity distance as
a function of redshift is given in the form
\begin{equation}
d_{L}(z)=\frac{1+z}{H_{0}}\int_{0}^{z}\frac{dz'}
{\sqrt{\Omega_{m,0}(1+z')^{3}-\Omega_{{\rm MOND},0}(1+z')^{2}\ln{(1+z')}
+(1-\Omega_{m,0})(1+z')^{2}}}.
\end{equation}
 
In the mentioned before paper by Lue and Starkman \cite{Lue:2003ky,Lue:2003if} 
we find very  interesting idea of the MOND law of gravitational interacting 
derived from the general relativity. The authors assuming the validity of the 
Birkhoff theorem and derive the basic cosmological model equation in the form
\begin{equation}
\frac{H^{2}}{H_{0}^{2}}=\frac{\dot{x}^{2}}{x^{2}}
=g\left(\frac{\rho}{\rho_{\rm crit}}\right)\equiv\left\{ \begin{array}{ll}
\Omega_{\text{m}}+C_{1}\Omega_{m}^{2/3} & \text{for the Einstein regime }
\Omega_{\text{m}}>\Omega_{c}\\
\beta\Omega_{m}^{2/3}\ln{\Omega_{\text{m}}}+C_{2}\Omega_{\text{m}}^{2/3} &
\textrm{for the MOND regime } \Omega_{\text{m}}<\Omega_{c}
\end{array} \right.
\end{equation}
 
In the Lue and Starkman \cite{Lue:2003ky,Lue:2003if} model the evolution of 
the universe consists of two phases, the first one dominated by gravity 
following the general relativity and the second one by its modification (MOGR). Our idea is
little different because we assume that both effects are acting as different
regimal effects but general
relativity dominates at early stages of evolution while MOGR describes the
late time evolution.  The dependence of the Hubble function describes
following  formula
\begin{equation}
H^{2}(x)=H_{0}^{2}\left\{\Omega_{\text{m},0}x^{-3}
+\Omega_{\text{MOND},0}x^{-2}\ln{x}+\Omega_{k,0}^{\text{eff}}x^{-2}\right\},
\end{equation}
where $\Omega_{k,0}^{\text{eff}}$ is the effective curvature such that
\begin{align}
&\Omega_{k,0}^{\text{eff}}=\beta\Omega_{\text{m},0}^{2/3}
\ln{\Omega_{\text{m},0}}+\Omega_{k,0},\\
&\Omega_{k,0}^{\text{eff}}=1-\Omega_{\text{m},0},\\
&\Omega_{\text{MOND},0}=-3\beta\Omega_{\text{m},0}^{2/3}.
\end{align}
Finally we obtain the same governing equation as (\ref{eq:14}) from the general
relativistic considerations.
 
The dynamics of this model can be represented in the form of the autonomous
dynamical system
\begin{equation}
\left\{ \begin{array}{ll}
\dot{x}=y \\
\dot{y}=-\partial V/\partial x
\end{array} \right.
\label{eq:20}
\end{equation}
where
\begin{equation}
V(x)=-\frac{1}{2}\left\{\Omega_{\text{m},0}x^{-1}
+\Omega_{\text{MOND},0}\ln{x}+\Omega_{k,0}^{\text{eff}}\right\}.
\end{equation}
The system (\ref{eq:20}) has the first integral in the form
\begin{center}
$\mathcal{H}=\frac{1}{2}y^{2}+V(x)=0$
\end{center}
 
From the first integral we obtain $d_{L}(z)$ relation
\begin{equation}
d_{L}(z)=\frac{1+z}{H_{0}}\int_{0}^{z}\frac{dz'}
{\sqrt{\Omega_{\text{m},0}(1+z')^{3}-\Omega_{\text{MOND},0}(1+z')^{2}\ln{(1+z')}
+\Omega_{k,0}^{\text{eff}}(1+z')^{2}}}.
\end{equation}
The parameter $\beta$ can be expressed as a function of $\Omega_{\text{m},0}$. 
For example for dust matter $\gamma=0$ we obtain constraint on $\beta$ parameter
\begin{equation}
\beta=-\frac{1-\Omega_{\text{m},0}}{3\Omega_{\text{m},0}^{2/3}\ln{\Omega_{\text{m},0}}}.
\end{equation}
If we define $z_{\text{eq}}$ as a moment in the evolution of the universe
at which both material and MOND terms are equal we obtain
\begin{equation}
\Omega_{\text{MOND},0}/\Omega_{\text{m},0}=\frac{1+z_{\text{eq}}}{\ln{(1+z_{\text{eq}})}},
\end{equation}
where $\Omega_{\text{MOND},0}/\Omega_{\text{m},0}=-3\beta\Omega_{\text{m},0}^{-1/3}$.
 
The results of our analysis are based on the Gold Riess Riess et al. \cite{Riess:2004nr})
supernovae Ia sample and there are presented in the table~\ref{results:1}.
On can see that considered models well fited SNIa data. However MOND model required
value of $\beta \simeq 15$ for possibility of explanation of flat rotation curve.
We obtained such value of $\beta$ only for the model with low $\Omega_{m,0}=0.01$
and $\Omega_{\mathrm{k},0}^{\mathrm{eff}}=-0.9$ (i.e $\Omega_{\mathrm{k},0}=-3.97$.
This value of $\Omega_{m,0}$ and $\Omega_{\mathrm{k},0}$ are in disagreement
with both results of CMBR and primordial  nucleosynthesis.
 
\begin{table}
\noindent
\caption{Results of the statistical analysis of the MOND model obtained
from the best fit with minimum $\chi^2$. F denotes fixed value of parameter.}
\label{results:1}
\begin{tabular}{@{}p{1.5cm}rrrrrrr}
\hline \hline
sample& $\Omega_{\mathrm{k},0}^{\mathrm{eff}}$  & $\Omega_{\mathrm{m},0}$ &
$\Omega_{\mathrm{MOND},0}$& $\Omega_{\Lambda,0}$ & $\mathcal{M}$ & $\chi^2$& $\beta$  \\
\hline
Gold & -0.88  &  0.00 &-2.00 & 1.88 &15.935&173.1 &$\infty$ \\
     &  ----  &  0.00 &-0.81 & 1.00 &15.955&175.2 &$\infty$ \\
     & -0.90  &  0.01F&-2.00 & 1.89 &15.935&173.1 &14.36    \\
     &  ----  &  0.01F&-0.78 & 0.99 &15.955&175.2 & 5.60    \\
     & -0.94  &  0.05F&-1.95 & 1.89 &15.935&173.1 & 4.78    \\
     &  ----  &  0.05F&-0.68 & 0.95 &15.955&175.3 & 1.67    \\
     & -0.83  &  0.30F&-1.19 & 1.53 &15.955&173.8 & 0.88    \\
     &  ----  &  0.30F&-0.02 & 0.70 &15.955&175.8 & 0.01    \\
     &  0.00  &  1.00 & 1.80 & ---- &15.965&177.6 &-0.60    \\
     &  ----  &  1.00 & 1.80 & ---- &15.965&177.6 &-0.60    \\
\hline
\end{tabular}
\end{table}

\section{MOFD model versus $\Lambda$CDM model in the light of Bayesian information criterion.}

In this section we extended previous model by adding dark energy in the form
cosmological constant or phantoms
We show that the MOFD cosmologies can be obtained as a Newtonian limit
of class of Phantom models which base on a simple modification of the FRW
equation. The physical status of both MOFD and Phantom models is similar
because they offer the possibility of alternative explanation of dark matter
and dark energy, respectively. We investigate some observational constraints
on the FRW cosmological models with baryonic matter and MOFD phase squeezed
in the evolutional scenario between the epoch of matter domination and the
dark energy epoch. We compare such a model with the concordance $\Lambda$CDM
model and argue that while both models are indistinguishable (close value
of $\chi^2$) the Akaike and Bayesian informative criterions favors MOFD model
with baryonic dark matter.
 
We consider two possible model with the exit on $\Lambda$ epoch or on the phantom
(Cardassian) epoch. For both cases the relations $H(z)$
are (respectively):
\begin{equation}
\label{eq:100}
H=H_0 \sqrt{\Omega_{\text{m},0}(1+z)^3+\Omega_{k \text{eff},0}(1+z)^2-\Omega_{\text{MOND},0}(1+z)^2ln(1+z) +\Omega_{\Lambda}}
\end{equation}
and
\begin{equation}
\label{eq:200}
H=H_0 \sqrt{\Omega_{\text{m},0}(1+z)^3+\Omega_{k \text{eff},0}(1+z)^2-\Omega_{\text{MOND},0}(1+z)^2ln(1+z) +\Omega_{Ph,0}(1+z)^{3n}}
\end{equation}
where $p=w\rho$, $w<-1$ for phantoms, $n=1+w$ for Cardassian.
 
\begin{sidewaystable}
\begin{tabular}{c|p{3.7cm}|llc}
\hline
case & name of model               & $\qquad H(z)$& free parameters & $d$ \\ \hline
0  & Einstein-de Sitter            & $H=H_0 \sqrt{\Omega_{\text{m},0}(1+z)^3+\Omega_{k,0}(1+z)^2}$ & $H_0,\Omega_{\text{m},0}$ & 2 \\
1  & $\Lambda$CDM                  & $H=H_0 \sqrt{\Omega_{\text{m},0}(1+z)^3+\Omega_{k,0}(1+z)^2+\Omega_{\Lambda}}$ & $H_0,\Omega_{\text{m},0},\Omega_{\Lambda}$ & 3 \\
2a & MOND, $\Omega_{m,0}$ - fitted & $H=H_0 \sqrt{\Omega_{\text{m},0}(1+z)^3+\Omega_{k \text{eff},0}(1+z)^2-\Omega_{\text{MOND},0}(1+z)^2ln(1+z) +\Omega_{\Lambda}}$ & $H_0,\Omega_{\text{m},0},\Omega_{\text{MOND},0},\Omega_{\Lambda}$ & 4 \\
2b & MOND, $\Omega_{\text{m},0}=0.05$     &     & $H_0,\Omega_{\text{MOND},0},\Omega_{\Lambda}$ & 3 \\
\hline
\end{tabular}
\caption{The Hubble function versus redshift for analyzed scenarios.}
\label{tab:1}
\end{sidewaystable}
 
\begin{sidewaystable}
\noindent
\label{tab:2}
\begin{tabular}{c|cccc}
\hline \hline
case & AIC (1-$\Omega_{\text{m},0}-\Omega_{\Lambda,0}=0$)
& AIC (1-$\Omega_{\text{m},0}-\Omega_{\Lambda,0} \ne 0$)
& BIC (1-$\Omega_{\text{m},0}-\Omega_{\Lambda,0}=0$)
& BIC (1-$\Omega_{\text{m},0}-\Omega_{\Lambda,0} \ne 0$)\\
\hline
0   & 325.5  & 194.4 & 328.6  & 200.5  \\
1   & 179.9  & 179.9 & 186.0  & 189.0  \\
2a  & 181.2  & 181.1 & 190.3  & 193.4  \\
2b  & 179.3  & 179.1 & 185.4  & 188.3  \\
\hline
\end{tabular}
\caption{The values of AIC and BIC for distinguished models
(Table \ref{tab:1}).}
\end{sidewaystable}
 
To compare considered models,how they fitted the data the informative criteria 
can be useful \cite{Liddle:2004nh}.
The problem of classification of the cosmological models on the light of
information criteria on the base of the astronomical data was discussed in
our previous papers
\cite{Godlowski05a,Szydlowski06a,Szydlowski06b,Szydlowski06c}.

The Akaike information criterion (AIC) is defined in the following way
\begin{equation}
\label{eq:111}
\text{AIC} = - 2\ln{\mathcal{L}} + 2d
\end{equation}
where $\mathcal{L}$ is the maximum likelihood and $d$ is the number of the
model parameters. The best model with a parameter set providing the preferred
fit to the data is that minimizes the AIC.
 
The Bayesian information criterion (BIC) introduced by Schwarz
is defined as
\begin{equation}
\label{eq:112}
\text{BIC} = - 2\ln{\mathcal{L}} + d\ln{N}
\end{equation}
where $N$ is the number of data points used in the fit.
 
This criterion gives a simple objective criterion for the inclusion of
new parameters into the standard $\Lambda$CDM model. From the results presented
in the Tables \ref{tab:1},\ref{tab:2} we can draw the following conclusion. 
The $\Omega_{MOND}$ is needed as a parameter and hence it is more likely that 
observations were generated in MOFD.

\section{Conclusions}
 
Our general conclusion is that MOND cosmology should be treated as a
potential alternative to the $\Lambda$CDM model in the context of explanation
of dark matter. To clarify the status of these model let us consider two sets
of best-fitted (gold sample of SNIa, Riess et al. \cite{Riess:2004nr})
model parameters (see table~\ref{results:1}).
 
Our point of view is following - because the $\Lambda$CDM model fits SNIa
data as well as the MOND alternative and additionally the second model explain
dark matter content in term of $\Omega_{\text{MOND},0}$ the model under
consideration should be treated as a possible candidate to explain dark
matter in the Universe.
However in particular case for the flat universe with
$\Omega_{\text{m},0}=0.3, \Omega_{\Lambda,0}=0.7$ we obtain $\Omega_{\text{MOND},0}=-0.02$,
i.e. $\beta=0.01$ while for the flat universe with
$\Omega_{\text{m},0}=0.05, \Omega_{\Lambda,0}=0.95$ we obtain $\Omega_{\text{MOND},0}=-0.68$
i.e. $\beta=1.67$.
The first case is corresponding to the $\Lambda$CDM model while the second
should be treated as an alternative description of acceleration driven by
cosmological constant and dumping by baryonic matter ($\Omega_{m,0}=0.05$).
Both models are indistinguishable---close values of $\chi^2$ (see
table~\ref{results:1}) and as result overlapping Hubble diagrams.
 
On can see that however considered models well fited SNIa data MOND model required
value of $\beta \simeq 15$ for possibility of explanation of flat rotation curve.
We obtained such value of $\beta$ only for the model with $\Omega_{m,0}=0.01$
and $\Omega_{\mathrm{k},0}^{\mathrm{eff}}=-0.9$ (i.e $\Omega_{\mathrm{k},0}=-3.97$).
This value of $\Omega_{m,0}$ and $\Omega_{\mathrm{k},0}$ are in disagreement both
with result of CMBR and early nucleosynthesis. Finally we conclude that, 
the MOND conception explain only separately  flat rotation curves of spiral 
galaxies or the fraction of dark matter in the Universe but it is not able to
explain these both facts together.
 
In this paper we also demonstrate that classical MOND conception can be
derived from more fundamental relativistic theory, namely from the generalized
Cardassian model.
 
The main aim of the paper was to show that the existence of the MOND phase
during the evolution of the Universe, before the epoch of domination of dark
energy can explain the presence of dark matter in the Universe. In other
words there are two indistinguishable scenario from the point of view of
explanation of the SNIa data. On the other hand if we {\'a} priori assume that
$\Omega_{\text{m},0} = 0.3$ the observations exclude the cosmological model
with the squeezing MOND phase in the cosmological scenario
($\Omega_{\text{MOND},0}  \simeq 0$). If we assume flat universe 
with the value of $\Omega_{\text{m},0} \simeq 0.3$
as it is suggested by extragalactic observations than we obtain that
$\Omega_{\text{MOND},0}$ should be small, but not necessary equal to zero.
In our approach we check whether the MOND phase frozen in the cosmological
scenario according to Starkman's idea can give us understanding of the
fraction of nonbarionic matter in $\Omega_{\text{m},0}$. We find that such a 
model well fit supernovae data but value of $\beta$ is far from Lue and 
Starkman \cite{Lue:2003if} value $\beta=15$

The second topic of this paper is the construction of the new class
of cosmological models with frozen the MOND phase into evolutional scenario
with exit to the Cardassian models. As it is well known the Cardassian models 
are an alternative to the cosmological models with dark energy in the
explanation of present acceleration of the current Universe. In these models
instead of dark energy violating the strong energy condition is
postulated a simple modification of the Friedmann first integral.
In this paper the model is fitted to observations of distant SNIa
using the Riess sample. We obtain analogous results as in the
case with the exit to the dark energy epoch. The advantage of
the model with frozen MOND phase and exit to the Cardassian models
is twofold. First, it can explain the acceleration of the Universe.
Second, it can explain the fraction of the dark matter.
 
The other results can be summarized as follows. We propose the theoretical 
description of cosmology MOFD based on the modified gravity. We find the 
connection of such models with recently discussed Cardassian models 
\cite{Godlowski-Szydlowski-Krawiec}. The parameter $\beta$ characterizing 
the MOND phase is estimated.  We also estimated this parameter
for the model with exit to the Cardassian model. In this case we obtain
the value of characteristic parameter $\beta$ which is far 
to the value assumed by Starkman ($\beta=15$). The value of Cardassian exponent
in the term $\rho^n$ in the modified Friedmann equation is close to zero.
This situation is very close the model with the cosmological constant
but nevertheless $n$ is negative and nonzero.
 
In this paper we pay attention to the flat cosmological models. It would
useful to make some remarks on the non-flat cosmological models.
In the models with exit to $\Lambda$ epoch we estimate the curvature type
term $\Omega_{k,0}^{\text{eff}}$. From the $\chi^2$ analysis we obtain that
non-flat case is more preferable than its flat counterpart. The similar
dependence of $d_{L}(z)$ on the Hubble diagram is obtained for fitting
the model with the exit to the Cardassian domination epoch with the SNIa
data (without any prior on $n$)

However MOFD cosmologies are compatible with observed late-time accelerated
expansion of contemporary universe. The popular method of apriorical
generalization of Friedmann equation is adding polytropic component of
r.h.s. of $H^2$ relation i.e. generalization Friedmann first equation.
Our proposal is generalization Raychaudhuri equation rather then Friedmann.
Then we obtain previous generalization plus one exceptional case which is 
strictly related with main subject of the paper.
 
however we still share the opinion expressed by Sahni that there is the 
fundamental difficulty of MOND gravity because this theory is not embedded 
within a more comprehensive and fundamental theory of gravitation. We also
do not know the Lagrangian for the Cardassian modification of gravity
but these models can be treated as a simple modification of the
cosmological models with FRW symmetry.

\section{Acknowledgments}
The work of M.S. was supported by project "COCOS" No. MTKD-CT-2004-517186.

\section{Appendix}
 
In this section we demonstrate how the presence of additional term in the Raychaudhuri equation can be modeled by some
noninteracting fictitious fluid $X$ with energy density $\rho_{X}(a)$ and pressure $p_{X}(a)$. We start from the basic
equations
\begin{subequations}
\label{eq:30}
\begin{align}
\frac{\ddot{a}}{a}&=-\frac{1}{6}(\rho+3p)+\frac{B}{6}a^{m}, \label{eq:30a} \\
\dot{\rho}&=-3H(\rho+p) \label{eq:30b}.
\end{align}
\end{subequations}
If we postulate that
\begin{equation}
-\frac{1}{6}(\rho_{X}+3p_{X})=\frac{B}{6}a^{m}
\label{eq:31}
\end{equation}
then (\ref{eq:30}) can be rewritten to the form
\begin{equation}
\frac{\ddot{a}}{a}=-\frac{1}{6}\sum_{i,X}(\rho_{k}+3p_{k}),
\label{eq:32}
\end{equation}
where the summation should be performed over all components of fluid.
For any $i$ fluid conservation equation is satisfied
\begin{equation}
\dot{\rho_{i}}=-3H(\rho_{i}+p_{i}).
\label{eq:33}
\end{equation}
Of course analogical condition should be satisfied by the fluid $X$, i.e
\begin{equation}
\frac{d\rho_{X}}{da}=-\frac{3}{a}(\rho_{X}+p_{X}).
\label{eq:34}
\end{equation}
From (\ref{eq:31}) we calculate $p_{X}$ and then we substitute this expression into (\ref{eq:34}). Hence we obtain
\begin{equation}
p_{X}=-\frac{1}{3}\rho_{X}-\frac{B}{3}a^{m}
\label{eq:35}
\end{equation}
and
\begin{equation}
\frac{d\rho_{X}(a)}{da}=-\frac{2}{a}\rho_{X}(a)+\frac{B}{a}a^{m}.
\label{eq:36}
\end{equation}
As a solution of (\ref{eq:36}) we obtain
\begin{equation}
\rho_{X}(a)=
\left\{ \begin{array}{ll}
\frac{C}{a^{2}}+\frac{B}{m+2}a^{m} & {\rm for}\quad m \neq -2 \\
\frac{C}{a^{2}}+\frac{B}{a^{2}}\ln{a} & {\rm for}\quad m = -2
\end{array} \right.
\label{eq:37}
\end{equation}
\begin{equation}
p_{X}(a)=
\left\{ \begin{array}{ll}
-\frac{C}{3a^{2}}-\frac{B}{3}\frac{m+3}{m+2}a^{m} & {\rm for}\quad m \neq -2 \\
-\frac{C+B}{3a^{2}}-\frac{B}{3a^{2}}\ln{a} & {\rm for}\quad m = -2
\end{array} \right.
\label{eq:37a}
\end{equation}
Of course system (\ref{eq:30}) has the first integral in the form
\begin{equation}
\rho_{\rm eff}-3\frac{\dot{a}^{2}}{a^{2}}=3\frac{k}{a^{2}}=\sum_{i}\left(\rho_{i}+\rho_{X}\right)-3\frac{\dot{a}^{2}}{a^{2}},
\label{eq:38}
\end{equation}
where
\begin{equation}
\dot{\rho_{i}}=-3H(\rho_{i}+p_{i})
\label{eq:39}
\end{equation}
for any $i$-fluid, $\sum_{i}\rho_{i}=\rho$ and also $\rho_{\rm eff}=-3H(\rho_{\rm eff}+p_{\rm eff})$.
The first integral (\ref{eq:38}) has different form for both distinguished cases
\begin{subequations}
\label{eq:40}
\begin{align}
\rho+\frac{C}{a^{2}}+\frac{B}{m+2}a^{m}-3\frac{\dot{a}^{2}}{a^{2}}=3\frac{k}{a^{2}},\quad {\rm for}\quad m \neq -2 \\
\rho+\frac{C}{a^{2}}+\frac{B}{a^{2}}\ln{a}-3\frac{\dot{a}}{a^{2}}=3\frac{k}{a^{2}},\quad {\rm for}\quad m=-2
\end{align}
\end{subequations}
We require the correspondence with standard FRW model for the case $B=0$.
Hence we obtain $C=0$. Finally the potential functions for both cases takes
the following form
\begin{equation}
V(a)=\left\{ \begin{array}{ll}
-\frac{1}{6}(\rho+\frac{B}{m+2}a^{m})a^{2} & {\rm for}\quad m \neq -2\\
-\frac{1}{6}(\rho+\frac{B}{a^{2}}\ln{a})a^{2} & {\rm for}\quad m = -2
\end{array} \right.
\label{eq:41}
\end{equation}
Of course the Hamiltonian system is still determined on the energy level $E=-k/2$ ($\mathcal{H}=\dot{a}^{2}/2+V(a)\equiv E$).


\begin{thebibliography}{8}
\expandafter\ifx\csname natexlab\endcsname\relax\def\natexlab#1{#1}\fi
\expandafter\ifx\csname bibnamefont\endcsname\relax
  \def\bibnamefont#1{#1}\fi
\expandafter\ifx\csname bibfnamefont\endcsname\relax
  \def\bibfnamefont#1{#1}\fi
\expandafter\ifx\csname citenamefont\endcsname\relax
  \def\citenamefont#1{#1}\fi
\expandafter\ifx\csname url\endcsname\relax
  \def\url#1{\texttt{#1}}\fi
\expandafter\ifx\csname urlprefix\endcsname\relax\def\urlprefix{URL }\fi
\providecommand{\bibinfo}[2]{#2}
\providecommand{\eprint}[2][]{\url{#2}}
 

\bibitem[{\citenamefont{Lue et~al.}(2004)\citenamefont{Lue, Scoccimarro, and
  Starkman}}]{Lue:2003ky}
\bibinfo{author}{\bibfnamefont{A.}~\bibnamefont{Lue}},
  \bibinfo{author}{\bibfnamefont{R.}~\bibnamefont{Scoccimarro}},
  \bibnamefont{and} \bibinfo{author}{\bibfnamefont{G.}~\bibnamefont{Starkman}},
  \bibinfo{journal}{Phys. Rev.} \textbf{\bibinfo{volume}{D69}},
  \bibinfo{pages}{044005} (\bibinfo{year}{2004}), \eprint{astro-ph/0307034}.
 
\bibitem[{\citenamefont{Lue and Starkman}(2004)}]{Lue:2003if}
\bibinfo{author}{\bibfnamefont{A.}~\bibnamefont{Lue}} \bibnamefont{and}
  \bibinfo{author}{\bibfnamefont{G.~D.} \bibnamefont{Starkman}},
  \bibinfo{journal}{Phys. Rev. Lett.} \textbf{\bibinfo{volume}{92}},
  \bibinfo{pages}{131102} (\bibinfo{year}{2004}), \eprint{astro-ph/0310005}.
 
\bibitem[{\citenamefont{Milgrom}(1983)}]{Milgrom:1983}
\bibinfo{author}{\bibfnamefont{M.}~\bibnamefont{Milgrom}},
  \bibinfo{journal}{Astrophys. J.} \textbf{\bibinfo{volume}{270}},
  \bibinfo{pages}{365} (\bibinfo{year}{1983}).
 
\bibitem[{\citenamefont{Freese and Lewis}(2002)}]{Freese:2002sq}
\bibinfo{author}{\bibfnamefont{K.}~\bibnamefont{Freese}} \bibnamefont{and}
  \bibinfo{author}{\bibfnamefont{M.}~\bibnamefont{Lewis}},
  \bibinfo{journal}{Phys. Lett.} \textbf{\bibinfo{volume}{B540}},
  \bibinfo{pages}{1} (\bibinfo{year}{2002}), \eprint{astro-ph/0201229}.
 
\bibitem[{\citenamefont{Perlmutter et~al.}(1999)}]{Perlmutter:1998np}
\bibinfo{author}{\bibfnamefont{S.}~\bibnamefont{Perlmutter}}
  \bibnamefont{et~al.} (\bibinfo{collaboration}{Supernova Cosmology Project}),
  \bibinfo{journal}{Astrophys. J.} \textbf{\bibinfo{volume}{517}},
  \bibinfo{pages}{565} (\bibinfo{year}{1999}), \eprint{astro-ph/9812133}.
 
\bibitem[{\citenamefont{Riess et~al.}(1998)}]{Riess:1998cb}
\bibinfo{author}{\bibfnamefont{A.~G.} \bibnamefont{Riess}} \bibnamefont{et~al.}
  (\bibinfo{collaboration}{Supernova Search Team}), \bibinfo{journal}{Astron.
  J.} \textbf{\bibinfo{volume}{116}}, \bibinfo{pages}{1009}
  (\bibinfo{year}{1998}), \eprint{astro-ph/9805201}.
 
\bibitem[{\citenamefont{Szydlowski and Czaja}(2004)}]{Szydlowski:2003cf}
\bibinfo{author}{\bibfnamefont{M.}~\bibnamefont{Szydlowski}} \bibnamefont{and}
  \bibinfo{author}{\bibfnamefont{W.}~\bibnamefont{Czaja}},
  \bibinfo{journal}{Phys. Rev.} \textbf{\bibinfo{volume}{D69}},
  \bibinfo{pages}{083518} (\bibinfo{year}{2004}), \eprint{gr-qc/0305033}.
 
\bibitem[{\citenamefont{Riess et~al.}(2004)}]{Riess:2004nr}
\bibinfo{author}{\bibfnamefont{A.~G.} \bibnamefont{Riess}} \bibnamefont{et~al.}
  (\bibinfo{collaboration}{Supernova Search Team}),
  \bibinfo{journal}{Astrophys. J.} \textbf{\bibinfo{volume}{607}},
  \bibinfo{pages}{665} (\bibinfo{year}{2004}), \eprint{astro-ph/0402512}.
 
\bibitem{Liddle:2004nh}
A.~R. Liddle, Mon. Not. Roy. Astron. Soc. 351 (2004) L49.
 
 
\bibitem{Godlowski05a}
W. Godlowski, M. Szydlowski, Phys. Lett. B 623 (2005) 10.
 
\bibitem{Szydlowski06a}
M. Szydlowski, W. Godlowski,  Phys. Lett. B 633 (2006) 427.
 
\bibitem{Szydlowski06b}
M. Szydlowski, W. Godlowski,  Phys. Lett. B 639 (2006) 5.
 
\bibitem{Szydlowski06c}
M. Szydlowski, A. Kurek, A. Krawiec, Phys.Lett. B 642 (2006) 171.
 
 
 
\bibitem{Akaike:1974}
H.~Akaike, IEEE Trans. Auto. Control 19 (1974) 716.
 
\bibitem{Schwarz:1978}
G.~Schwarz, Annals of Statistics 5 (1978) 461.
 
\bibitem{Bennett:2003bz}
C.~L. Bennett, et~al., Astrophys. J. Suppl. 148 (2003) 1.
 
\bibitem{Jeffreys:1961}
H.~Jeffreys, Theory of Probability, 3rd Edition, Oxford University Press,
  Oxford, 1961.
 
\bibitem{Mukherjee:1998wp}
S.~Mukherjee, E.~D. Feigelson, G.~J. Babu, F.~Murtagh, C.~Fraley, A.~Raftery,
  Astrophys. J. 508 (1998) 314.
 
\bibitem{Peebles:2002gy}
P.~J.~E. Peebles, B.~Ratra, Rev.   Mod. Phys. 75 (2003) 559.
 
\bibitem{Godlowski-Szydlowski-Krawiec}
W.~Godlowski, M.~Szydowski, A. Krawiec, APJ  605 (2004) 599.
 
\end{thebibliography}

\end{document}